\documentclass[11pt]{article}
\usepackage{amsfonts, amsmath, amssymb, amsthm, bbm, enumerate, float, graphicx, latexsym, mathrsfs, pgf,  setspace, stmaryrd, tikz, verbatim, wrapfig, xparse}

\usetikzlibrary{calc}

\usepackage{ifthen}

\usepackage[margin = 1 in]{geometry}

\theoremstyle{definition}
	\newtheorem{defn}{Definition}[section]
\theoremstyle{definition}

\theoremstyle{plain}
	\newtheorem{thm}[defn]{Theorem}
	\newtheorem{lem}[defn]{Lemma}
	
	\newtheorem{prop}[defn]{Proposition}
\theoremstyle{remark}

\newcommand{\Q}{\mathbb{Q}}
\newcommand{\C}{\mathbb{C}}

\DeclareMathOperator{\Sp}{Sp}

\newcommand{~}{\sim}

\NewDocumentCommand\eigenA{mg}{%
    \ensuremath{E_{#1}{\IfNoValueTF{#2}{}{(#2)}}}%
}
\NewDocumentCommand\eigenL{mg}{%
    \ensuremath{F_{#1}{\IfNoValueTF{#2}{}{(#2)}}}%
}

\newcommand{\emptyg}[1]{\overline{K}_{#1}}
\newcommand{\norm}[1]{\left\lVert #1 \right\rVert}

\newcommand{\usedirac}{}

\ifthenelse{\isundefined{\usedirac}}
{ 
	\newcommand{\ket}[1]{\mathbf{#1}}
	\newcommand{\bra}[1]{\mathbf{#1}^{\dagger}}
	\newcommand{\eket}[1]{{\mathbf{e}}_{#1}}	
	\newcommand{\ebra}[1]{\eket{#1}^{T}}      	
	\newcommand{\ekettuple}[1]{{\mathbf{e}}_{(#1)}}			
	\newcommand{\ebratuple}[1]{\ekettuple{#1}^{T}}      	
	\newcommand{\outprod}[2]{\ket{#1}\bra{#2}}				
	\newcommand{\eoutprod}[2]{\eket{#1}\ebra{#2}}						
	\newcommand{\allzeroket}[1]{\mathbf{0}_{#1}}
	\newcommand{\allzerobra}[1]{\mathbf{0}^{T}_{#1}}
	\newcommand{\alloneket}[1]{\mathbf{j}_{#1}}
	\newcommand{\allonebra}[1]{\mathbf{j}^{T}_{#1}}
}
{ 
	\newcommand{\bra}[1]{\left\langle #1 \right|}
	\newcommand{\ket}[1]{\left| #1 \right\rangle}
	\newcommand{\ebra}[1]{\bra{#1}}
	\newcommand{\eket}[1]{\ket{#1}}
	\newcommand{\ebratuple}[1]{\bra{#1}}
	\newcommand{\ekettuple}[1]{\ket{#1}}
	
	\newcommand{\outprod}[2]{\ket{#1}\! \bra{#2}}
	
	\newcommand{\eoutprod}[2]{\outprod{#1}{#2}}
	\newcommand{\allzeroket}[1]{\ket{0_{#1}}}
	\newcommand{\allzerobra}[1]{\bra{0_{#1}}}
	\newcommand{\alloneket}[1]{\ket{j_{#1}}}
	\newcommand{\allonebra}[1]{\bra{j_{#1}}}
}

\newcommand{\twovector}[2]{\left[\begin{array}{c} #1 \\ #2 \end{array}\right]}

\newcommand{\corona}{\circ}

\newcommand{\ignore}[1]{}
\newcommand{\etal}{{\it et al. }}

\newcommand*{\boo}[1]{\ensuremath\overrightarrow{#1}}
\newcommand{\hseq}{\boo{H}}

\newif\ifnotesw\noteswtrue
   {\ifnotesw\marginpar[\hfill\(\top\)]{\(\top\)}\fi}%
   {\ifnotesw\marginpar[\hfill\(\bot\)]{\(\bot\)}\fi}

\newcommand{\mnote}[1]%
    {\ifnotesw\marginpar%
        [{\scriptsize\begin{minipage}[t]{\marginparwidth}
        \raggedleft#1%
                        \end{minipage}}]%
        {\scriptsize\begin{minipage}[t]{\marginparwidth}
        \raggedright#1%
                        \end{minipage}}%
    \fi}

\begin{document}

\title{Laplacian State Transfer in Coronas}
\author{
Ethan Ackelsberg\thanks{Division of Science, Mathematics and Computing, Bard College at Simon's Rock.}
\and
Zachary Brehm\thanks{Department of Mathematics, SUNY Potsdam.}
\and
Ada Chan\thanks{Department of Mathematics and Statistics, York University.}
\and
Joshua Mundinger\thanks{Department of Mathematics and Statistics, Swarthmore College.}
\and
Christino Tamon\thanks{Department of Computer Science, Clarkson University.}}
\maketitle

\begin{abstract}
We prove that the corona product of two graphs has no Laplacian perfect state transfer
whenever the first graph has at least two vertices.
This complements a result of Coutinho and Liu who showed that no tree of size greater than two
has Laplacian perfect state transfer.
In contrast, we prove that the corona product of two graphs exhibits 
Laplacian pretty good state transfer, under some mild conditions.
This provides the first known examples of families of graphs with Laplacian pretty good state transfer.
Our result extends of the work of Fan and Godsil on double stars to the Laplacian setting.
Moreover, we also show that the corona product of any cocktail party graph with a single vertex graph
has Laplacian pretty good state transfer, even though odd cocktail party graphs 
have no perfect state transfer.
\end{abstract}

\section{Introduction}

Given a graph $G$ and a symmetric matrix $M$ associated with $G$, the continuous-time quantum walk on $G$ 
relative to $M$ is given by the unitary matrix 
\begin{equation}
U(t) := \exp(-itM).
\end{equation}
This notion was introduced by Farhi and Gutmann \cite{fg98} as a paradigm to design efficient quantum algorithms.
Physically, this also represents the evolution of a quantum spin system. 
This interesting connection was explored in the works of Bose \cite{b03} and Christandl \etal \cite{cdel04,cddekl05}.
Furthermore, as pointed out by Bose \etal \cite{bcms09}, 
there are two different matrices $M$ of interest. In the so-called XY model, $M$ is the adjacency matrix of $G$;
in the XYZ model, $M$ is the Laplacian of $G$.
For details on these physical models, see Bose \etal \cite{bcms09} for a derivation. 
Note that if $G$ is regular, these quantum walks differ only by complex conjugation and a phase factor.

From the physical standpoint, quantum walks relative to the adjacency matrix and the Laplacian are equally important.
However, the current literature has focused mostly on quantum walks relative to the adjacency matrix.
In this paper, we investigate continuous-time quantum walks relative to the Laplacian.

We are interested in the phenomenon of state transfer, which models the routing of information between particles 
in the associated spin system.
This was the original motivation of the work by Bose \cite{b03}.
A graph $G$ is said to have perfect state transfer between vertices $u$ and $v$ if there is a time $t$ such that
\begin{equation}
	\left| \exp(-itM)_{uv}\right|^2 = 1.
\end{equation}
Physically, this means that the probability of state transfer between vertices $u$ and $v$ is unity.
We will refer to the matrix entry $\exp(-itM)_{uv}$ as the transition element between the vertices $u$ and $v$.

There are several infinite families of graphs known to have perfect state transfer.
This includes hypercubes \cite{cddekl05}, some families of distance-regular graphs \cite{cggv15}, 
complete graphs with a missing edge \cite{bcms09}, and some joins \cite{normalized14}. 
However, recently it has become clear that perfect state transfer is rare. 
In the adjacency matrix case, Godsil showed that there are at most finitely many graphs with 
a given maximum valency with perfect state transfer \cite{g12}, while, in the Laplacian case, 
Coutinho and Liu showed that there is no perfect state transfer on trees with at least three vertices \cite{cl14}.

Nevertheless, transmission of information in a quantum system may not occur perfectly, 
but rather with probability that is arbitrarily close to unity. 
We thus consider a relaxation. A graph $G$ is said to have pretty good state transfer if for each $\epsilon > 0$, there exists a time $t$ such that
\begin{equation}
		\left|\exp(-itM)_{uv}\right|^2 \geq 1 - \epsilon.
\end{equation}
This notion was proposed by Godsil \cite{g11-survey}.
Relative to the adjacency matrix, pretty good state transfer was studied on paths \cite{gkss12} and double stars \cite{fg13}. 
However, prior to this work, there are no known families of graphs with pretty good state transfer 
in Laplacian quantum walks. 
In this paper, we provide the first infinite families of graphs with Laplacian pretty good state transfer.
Our families of graphs are constructed using the corona product of two graphs.

The corona product was introduced by Frucht and Harary to construct a graph whose automorphism group is the wreath product of the automorphism groups of the component graphs \cite{fh70}.
Quantum walks (in a discrete-time setting) on the corona of hypercubes and cliques were studied numerically 
by Makmal \etal \cite{mzmtb14}. 
In this work, we show that there is no perfect state transfer on coronas, continuing to support that perfect state transfer is rare.
However, we show that given a graph $G$ with perfect state transfer, the corona product of $G$ with another graph $H$ will have pretty good state transfer, subject to a condition on the number of vertices of $H$. 
As a corollary,
we extend the work of Fan and Godsil on double stars \cite{fg13} to the Laplacian, as well as construct an infinite family of coronas with pretty good state transfer where neither component of the product has any state transfer.

For more information on algebraic graph theory, see Godsil and Royle \cite{gr01},
and for a survey on state transfer on graphs, see Godsil \cite{g11-survey}.

\section{Preliminaries}

Let $G$ be a graph with adjacency matrix $A$ and diagonal degree matrix $D$. 
The \textit{Laplacian} of $G$ is defined as the matrix $L := D - A$. The unitary matrix

\begin{equation}
	U(t) := \exp(-itL)
\end{equation}
determines a \textit{continuous-time Laplacian quantum walk} on $G$. 
We are interested in studying {perfect state transfer} and {pretty good state transfer}
in such a quantum walk.
If $G$ has $n$ vertices,
we associate the vertices of $G$ with coordinates in $\C^n$, 
and let $\eket{u}$ denote the characteristic vector of the vertex $u$.

A graph $G$ has \emph{perfect state transfer} between vertices $u$ and $v$ at time $\tau$ if there exists a complex number 
$\gamma$ such that
\begin{equation} \label{eq: pst-definition}
	U(\tau) \eket{u} = \gamma \eket{v}.
\end{equation}
The complex number $\gamma$ is called the \textit{phase} of the perfect state transfer.
Since $U(t)$ is unitary for all $t$, the condition \eqref{eq: pst-definition} is equivalent to
$|U(\tau)_{uv}|^2 = 1$. 
Since $L$ is real symmetric, $U(t)$ is symmetric, which shows that \eqref{eq: pst-definition} is symmetric in $u$ and $v$,
that is, $U(\tau)\eket{v} = \gamma\eket{u}$. 

\ignore{
Since $L$ is symmetric, the Spectral Theorem shows that $L$ admits a spectral decomposition.

\begin{thm}[Spectral Theorem]
    For any symmetric matrix $M$ with eigenvalues $0 = \lambda_0 <  \lambda_1 <  \ldots < \lambda_d$, 
	there exist Hermitian projection matrices $\eigenL{\lambda_0}, \eigenL{\lambda_1}, \ldots, \eigenL{\lambda_d}$ such that
	\begin{enumerate}[i)]
		\item $\eigenL{\lambda_i}\eigenL{\lambda_j} = \delta_{ij}\eigenL{\lambda_i}$
		\item $\sum_{i=0}^d \eigenL{\lambda_i} = I$
		\item $\sum_{i=0}^d \lambda_i \eigenL{\lambda_i} = M$
	\end{enumerate}
\end{thm}
}

By the spectral theorem, $L$ admits a decomposition
\begin{equation}
L= \sum_{i=0}^d \lambda_i \eigenL{\lambda_i},
\end{equation}
where $0 = \lambda_0 <  \lambda_1 <  \ldots < \lambda_d$ are the distinct eigenvalues of $L$ 
and $\eigenL{\lambda_i}$ is the eigenprojector for eigenvalue $\lambda_i$.
Note that $\sum_{i=0}^d \eigenL{\lambda_i} = I$.

We say two vertices $u$ and $v$ are \textit{strongly cospectral} if, for every eigenvalue $\lambda$ of $G$, we have 
\begin{equation}
	\eigenL\lambda \eket{u} = \pm \eigenL\lambda \eket{v}.
\end{equation}
The \textit{eigenvalue support} of a vertex $v$ is the set of all eigenvalues $\lambda$ of $G$ 
such that $\eigenL\lambda \eket{v} \neq 0$.

The following theorem states the known necessary and sufficient conditions for Laplacian perfect state transfer.

\begin{thm}[Coutinho \cite{c14}, Theorem 7.3.1] \label{thm: coutinho-pst-characterization}
	Let $G$ be a graph, and let $u$ and $v$ be vertices in $G$. 
	Let $S$ be the eigenvalue support of $u$.
	Then, there is Laplacian perfect state transfer between $u$ and $v$
	at time $\tau$ 
	if and only if all of the following hold:
	\begin{enumerate}[(i)]
		\item The vertices $u$ and $v$ are strongly cospectral;
		\item All eigenvalues in $S$ are integers;
		\item For each $\lambda \in S$, $\ebra{u}\eigenL{\lambda}\eket{v}$ is positive 
			if and only if $\lambda/\gcd(S)$ is even.
	\end{enumerate}
	Moreover, if these hold, there is a minimum time of perfect state transfer given by $t_0 := \pi/\gcd(S)$, 
	and $\tau$ is an odd multiple of $t_0$. 	
\end{thm}

A graph $G$ has \textit{pretty good state transfer} between vertices $u$ and $v$ if 
for every $\epsilon > 0$, there exists a time $t$ and there exists a complex number $\gamma$
such that 
\begin{equation}
	\norm{U(t) \eket{u} - \gamma \eket{v}} < \epsilon.
\end{equation}
Since $U(t)$ is unitary, this is equivalent to requiring that for each $\epsilon > 0$, there exists a time $t$ such that
\begin{equation}
	|U(t)_{uv}|^2 > 1 - \epsilon.
\end{equation}

\medskip
As in Godsil \etal \cite{gkss12},
our main tool for showing the existence of pretty good state transfer is Kronecker's Approximation Theorem. 

\begin{thm}[Hardy and Wright \cite{hw00}, Theorem 442] \label{thm: kronecker} 
Let $1,\lambda_{1},\ldots,\lambda_{m}$ be linearly independent over $\Q$.
Let $\alpha_{1},\ldots,\alpha_{m}$ be arbitrary real numbers,
and let $\epsilon$ be a positive real number.
Then, there is an integer $\ell$ and integers $q_1,\ldots, q_m$
so that
\begin{equation}\label{eq: kroneckers-theorem}
|\ell\lambda_{k} - \alpha_{k} - q_{k}| < \epsilon,
\end{equation}
for each $k=1,\ldots,m$.
\end{thm}

For brevity, whenever we have an equation of the form $|\alpha - \beta| < \epsilon$,
we will write instead $\alpha \approx \beta$ and omit the explicit dependence on
$\epsilon$. For example, \eqref{eq: kroneckers-theorem} will be represented as $\ell\lambda_k - q_k \approx \alpha_k$.

In our applications of Kronecker's Theorem, we will use the following lemma to identify sets of numbers which are
linearly independent over the rationals.

\begin{lem}[Richards \cite{r74}] \label{lem: square-free-root-independence} 
The set $\{\sqrt{\Delta} : \Delta \, \text{is a square-free integer} \}$ is linearly independent over the 
set of rational numbers $\Q$.
\end{lem}

\par\noindent
\textit{Notation:}
Let $\alloneket{m}$ denote the all-one vector with $m$ components, and let $J_{m,n}$ denote the $m \times n$ all-one matrix
(or simply $J_{m}$ if $m = n$).
In cases where we need to specify the underlying graph, we will use a notation such as $L(G)$ instead of $L$.
Throughout this paper, the spectrum of $G$ is the spectrum of its Laplacian $L$, denoted $\Sp(G)$, unless stated otherwise.

\section{Corona of Graphs}

Let $G$ be a graph on the vertex set $V(G) = \{v_1, \ldots, v_n\}$ and let $\hseq = (H_1, \dots, H_n)$
be an $n$-tuple of graphs.
The \emph{(inhomogenous) corona} $G \corona \hseq$ is formed by taking the disjoint union of $G$ 
and $H_1, \dots, H_n$ and then adding an edge from each vertex in $H_j$ to the vertex $v_j$ in $G$. 
Formally, the corona $G \corona \hseq$ has the vertex set
	\begin{equation}
		V(G \corona \hseq) = \{ (v, 0) : v \in V(G) \} \cup 
			\bigcup_{j=1}^{n} \{ (v_j, w) : v_j \in V(G), w \in V(H_j) \}
	\end{equation}
	and the adjacency relation
	\begin{equation}
		(v_j, w) ~ (v_k, w') \iff
		 \begin{cases}
		 	w = w' = 0 \, \ \text{and} \, \ v_j ~_G v_k, & \text{or} \\
			v_j = v_k \, \ \text{and} \, \ w ~_{H_j} w', & \text{or} \\
			v_j = v_k \, \ \text{and exactly one of} \, \ w \, \ \text{and} \, \ w' \, \ \text{is} \, \ 0.
		\end{cases}
	\end{equation}

\par\noindent
For notational convenience, we will identify $V(G)$ with $\{1,2,\ldots,n\}$,
whereby we simply denote $v_{\ell}$ by $\ell$.

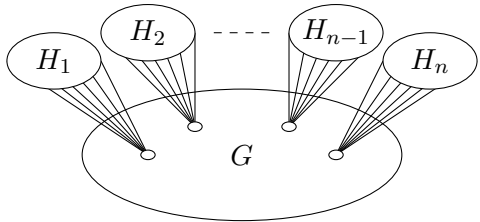
\begin{figure}[H]
	\centering
	\begin{tikzpicture}[scale=1.25]
		\draw (0,0) node{$G$} circle[x radius = 1.7, y radius = 0.7];
		\foreach \x in {1,2,4,5} {
			\foreach \y in {0,...,5}
				\draw (-1.5+0.5*\x, -0.1*\x*\x + 0.6*\x - 0.5) -- ++(-2+0.5*\x+0.2*\y, 1);
			\filldraw[fill=white] (-1.5+0.5*\x, -0.1*\x*\x + 0.6*\x - 0.5) circle[x radius = 0.03in, y radius = 0.02in] node{};
		}
		\filldraw[fill=white] (-2, 1) circle[x radius=0.5, y radius=0.3] node{$H_1$};
		\filldraw[fill=white] (-1, 1.3) circle[x radius=0.5, y radius=0.3] node{$H_2$};
		\draw[dashed] (-0.3,1.3) -- ++(0.6,0);
		\filldraw[fill=white] (1, 1.3) circle[x radius=0.5, y radius=0.3] node{$H_{n-1}$};
		\filldraw[fill=white] (2, 1) circle[x radius=0.5, y radius=0.3] node{$H_n$};
	\end{tikzpicture}
	\caption{An inhomogeneous corona $G \corona \protect \hseq$ where $\protect \hseq = (H_{1},\ldots,H_{n})$.}
\end{figure}

The adjacency matrix of the corona $G \corona \hseq$ is given by
\begin{equation}
	A(G \corona \hseq) = A(G) \otimes \eoutprod{0}{0} 
	 + \sum_{\ell=1}^n \eoutprod{\ell}{\ell} \otimes
		\left[\begin{array}{cc}
			0 & \allonebra{m} \\
			\alloneket{m} & A(H_\ell)
		\end{array}\right].
\end{equation}
Then, it follows that the Laplacian of $G \corona \hseq$ is
\begin{equation} \label{eq: corona-laplacian-matrix}
	L(G \corona \hseq) 
	= \left(L(G) + mI\right) \otimes \eoutprod{0}{0} 
	+ \sum_{\ell=1}^n \eoutprod{\ell}{\ell} \otimes
		\left[\begin{array}{cc}
			0 & -\allonebra{m} \\
			-\alloneket{m} & L(H_\ell) + I
		\end{array}\right].
\end{equation}

The definition given by Frucht and Harary \cite{fh70} coincides with when $\hseq$ is a constant sequence $(H, \ldots, H)$ for some graph $H$. We will denote such a corona with a constant sequence simply as $G \corona H$. 
In this case, the spectrum of $G \corona H$ is known. 

\begin{thm}[Barik \etal \cite{bps07}, Theorem 3.2] \label{thm: homogenous-corona-spectrum}
	Let $G$ be a graph on $n$ vertices and $H$ be a graph on $m$ vertices. 
	Suppose $G$ has spectrum $0 = \lambda_0 < \ldots < \lambda_p$ with multiplicities $r_0, \ldots, r_p$, 
	and $H$ has spectrum $0 = \mu_0 < \mu _1 < \ldots < \mu_q$ with multiplicities $r'_0, \ldots, r'_q$. 
	Then the (homogeneous) corona $G \corona H$ has the following spectrum:
	\begin{enumerate}[(a)]
		\item $1$ with multiplicity $n(r'_0 - 1)$;
		\item $\mu_j + 1$ with multiplicity $nr'_j$, for each $j = 1,\ldots,q$;
		\item $\frac{1}{2}\left( \lambda_j + m + 1 \pm \sqrt{(\lambda_j + m + 1)^2 - 4\lambda_j}\right)$
			with multiplicity $r_j$, for each $j = 1,\ldots,p$.
	\end{enumerate}
\end{thm}

In the following, 
we extend Theorem \ref{thm: homogenous-corona-spectrum} by computing the eigenvalues and eigenprojectors of 
the inhomogenous corona $G \corona (H_{1},\ldots,H_{n})$ 
when the order $|V(H_{j})|$ is the same for each $j = 1,\ldots,n$. 

\begin{prop} \label{prop: lap-spec-decomp-corona}
Let $G$ be a graph on $n$ vertices and $\hseq = (H_{1},\ldots,H_{n})$ be an $n$-tuple of graphs on $m \geq 1$ vertices.
Then the spectrum of $G \corona \hseq$ is given by:
\begin{enumerate}[(a)]
\item
	$1$ is an eigenvalue, if there exists $H_{\ell}$ that is disconnected,
	with the eigenprojector
	\begin{equation} \label{eq: corona-1-eigenprojector}
	\eigenL{1} = 
		\sum_{\ell = 1}^{n} \eoutprod{\ell}{\ell} \otimes 
			\left[\begin{array}{cc}
				0 & \allzerobra{m} \\
				\allzeroket{m} & \eigenL{0}{H_\ell} - \frac{1}{m}J_m
			\end{array}\right].
	\end{equation}
	Note $\eigenL{0}{H_{j}} = \frac{1}{m}J_{m}$ if and only if $H_{j}$ is connected.

\item 
	$\mu+1$ is an eigenvalue, if there exists $H_{\ell}$ that has a nonzero eigenvalue $\mu$,
	with the eigenprojector 
	\begin{equation} \label{eq: corona-mu-eigenprojector}
	\eigenL{\mu+1} = 
		\sum_{\ell = 1}^n \eoutprod{\ell}{\ell} \otimes 
			\left[\begin{array}{cc}
				0 & \allzerobra{m} \\
				\allzeroket{m} & \eigenL{\mu}{H_\ell}
			\end{array}\right],
	\end{equation}
	where we assume $F_\mu(H_\ell) = 0$ if $\mu$ is not an eigenvalue of $H_\ell$.

\item 
	$\lambda_\pm = \frac{1}{2}(m + \lambda + 1 \pm \sqrt{(m + \lambda - 1)^2 + 4m})$
	are eigenvalues, for each eigenvalue $\lambda$ of $G$,
	with eigenprojectors
	\begin{equation} \label{eq: corona-pm-eigenprojector}
	\eigenL{\lambda_\pm} = 
		\frac{1}{(1- \lambda_\pm)^2 + m} \eigenL{\lambda}{G} \otimes 
			\left[\begin{array}{cc}
				(1- \lambda_\pm )^2 		& (1 - \lambda_\pm)\allonebra{m} \\
				(1 - \lambda_\pm)\alloneket{m}	& J_m
			\end{array}\right].
	\end{equation}
\end{enumerate}
Therefore, the spectral decomposition of the Laplacian of the corona $G \corona \hseq$ is given by
\begin{equation}
L(G \corona \hseq) = 
	\sum_{\lambda \in \Sp(G)} \sum_{\pm} \lambda_{\pm} \eigenL{\lambda_\pm}
	+
	\sum_{\mu} (\mu+1) \eigenL{\mu+1},
\end{equation}
where the sum over $\mu$ is over all eigenvalues of the graphs $H_\ell$, for $\ell=1,\ldots,n$.

\begin{proof}
Let $H_{\ell}$ be one of the graphs in the sequence $\hseq$.
Suppose $\ket{x}$ is a normalized eigenvector of $L(H_\ell)$ corresponding to eigenvalue $\mu$,
and that $\ket{x}$ is orthogonal to $\alloneket{m}$. Then,
\begin{equation}
	L(G \corona \hseq) \ \eket{\ell} \otimes \twovector{0}{\ket{x}} = (\mu+1) \ \eket{\ell} \otimes \twovector{0}{\ket{x}}.
\end{equation}
If $H_\ell$ is disconnected, then $L(H_\ell)$ has eigenvectors with eigenvalue zero orthogonal to $\alloneket{m}$.
In this case, $1$ is an eigenvalue of $L(G \corona \hseq)$ with the eigenprojector given in
\eqref{eq: corona-1-eigenprojector}.
For each eigenvalue $\mu \neq 0$ of $L(H_\ell)$ with eigenprojector $\eigenL{\mu}{H_\ell}$, 
$\mu + 1$ is an eigenvalue of $L(G \corona \hseq)$ with the eigenprojector given in
\eqref{eq: corona-mu-eigenprojector}.

Since $|V(G)| = n$, then \eqref{eq: corona-1-eigenprojector} and \eqref{eq: corona-mu-eigenprojector}
together give $n(m-1)$ eigenvectors of $L(G \corona \hseq)$.
We construct the remaining $2n$ eigenvectors using the eigenvectors of $L(G)$.
Suppose $\ket{y}$ is an eigenvector of $L(G)$ with eigenvalue $\lambda$. Then,
\begin{align} \label{eq: nonHeigenvecs}
	L(G \corona \hseq) \ \ket{y} \otimes \twovector{1-\lambda_{\pm}}{\alloneket{m}}
	 = \lambda_\pm \ \ket{y} \otimes \twovector{1-\lambda_{\pm}}{\alloneket{m}}
\end{align}
if and only if 
\begin{align}
	\lambda_\pm 
        = \frac{m + \lambda + 1 \pm \sqrt{(m + \lambda + 1)^2 - 4\lambda}}{2}
		= \frac{m + \lambda + 1 \pm \sqrt{(m + \lambda - 1)^2 + 4m}}{2}.
\end{align}
After normalizing, the eigenprojectors corresponding to eigenvalues $\lambda_\pm$ are given by \eqref{eq: corona-pm-eigenprojector}.
\end{proof}
\end{prop}

\par\noindent
Note that when $\hseq=(H,\ldots,H)$, Proposition \ref{prop: lap-spec-decomp-corona} coincides with 
Theorem \ref{thm: homogenous-corona-spectrum}.

\section{Perfect State Transfer}

In the following theorem, we show that there is no perfect state transfer between any pair of vertices in
the corona of two graphs whenever the first graph has at least two vertices. 
The proof relies on the fact that there are no integer eigenvalues in the support of any vertex.

\begin{thm}	\label{thm: nolaplacianPST}
	Let $G$ be a connected graph on $n \ge 2$ vertices and $\hseq = (H_1, \dots, H_n)$ 
	be an $n$-tuple of graphs on $m \ge 1$ vertices. 
	Then there is no Laplacian perfect state transfer in $G \corona \hseq$.
	\begin{proof}
		Let $(v,w)$ be a vertex of $G \corona \hseq$. 
		By Theorem \ref{thm: coutinho-pst-characterization},
		it suffices to find a non-integer eigenvalue in the support of $(v,w)$.
		Since $G$ is connected on at least two vertices, 
		there exists a positive eigenvalue $\lambda$ in the eigenvalue support of $v$.
		From Proposition \ref{prop: lap-spec-decomp-corona}, both 
		\begin{equation}
			\lambda_\pm = \frac{1}{2}\left( m + \lambda + 1 \pm \sqrt{( m + \lambda - 1)^2 + 4m}\right)
		\end{equation}		
		are in the eigenvalue support of $(v,w)$. 
		Suppose towards contradiction that both $\lambda_\pm$ are integers. Then both
		\begin{align}
			m + \lambda + 1 = \lambda_+ + \lambda_-, \\
			\sqrt{(m + \lambda - 1)^2 + 4m } = \lambda_+ - \lambda_-
		\end{align}
		are integers, implying that $\lambda$ is an integer.
		Further, $(m + \lambda - 1)^2 + 4m$ is a perfect square, but since $4m$ is even,
		the parity of this square must be the same as $(m + \lambda - 1)^2$. 
		Since $\lambda > 0$, the least square greater than $(m+\lambda - 1)^2$ 
		with the same parity is $(m + \lambda + 1)^2$. This yields the bound
		\begin{align}
			(m + \lambda - 1)^2 + 4m \ge (m + \lambda + 1)^2
		\end{align}
		which implies $\lambda \leq 0$, a contradiction. Thus, one of $\lambda_\pm$ is not an integer.
	\end{proof}
\end{thm}

\section{Pretty Good State Transfer}

Although Theorem \ref{thm: nolaplacianPST} shows there is no perfect state transfer on coronas, 
we will show that there is pretty good state transfer, under some mild conditions.
First, we state a useful form for the transition elements of a Laplacian quantum walk on coronas.

\begin{prop} \label{prop: lap-transition-element}
	Let $G$ be a graph on $n$ vertices and let $\hseq = (H_{1},\ldots,H_{n})$ be an $n$-tuple of graphs on $m\geq 1$ vertices.
	If $u$ and $v$ are vertices of $G$, 
	then the transition element between vertices $(u,0)$ and $(v,0)$ in $G \corona \hseq$ is given by
	\begin{multline} \label{eq: Lmatrixelement}
		\ebratuple{u,0} \exp(-itL(G\corona \hseq)) \ekettuple{v,0} \\
		= e^{-it (m+1)/2} 
			\sum_{\lambda \in \Sp(G)} e^{-it\lambda/2} 
			\ebra{u} \eigenL{\lambda}{G} \eket{v}
			\left( \cos\left(\frac{t}{2}\Delta_\lambda \right) - 
			\frac{(m + \lambda - 1)}{\Delta_\lambda} i\sin\left(\frac{t}{2}\Delta_\lambda \right)\right),
	\end{multline}
	where $\Delta_\lambda := \sqrt{ (m + \lambda - 1)^2 + 4m}$, for each eigenvalue $\lambda$ of $G$.
\begin{proof}
	For each eigenvalue $\lambda$ of $L(G)$, recall
	\begin{align}
		\lambda_{\pm} = \frac{1}{2}\left(m+\lambda+1 \pm \Delta_\lambda\right).
	\end{align}
	By Proposition \ref{prop: lap-spec-decomp-corona}, the transition element between vertices $(u,0)$ and $(v,0)$ is given by
	\begin{equation}
	\ebratuple{u,0} e^{-itL(G \corona H)} \ekettuple{v,0}
	=
	\sum_{\lambda \in \Sp(G)} e^{-i(m+\lambda+1)t/2}
		\ebra{u}\eigenL{\lambda}{G}\eket{v}
		\left(\sum_{\pm} e^{\mp i\Delta_{\lambda}t/2} 
		\left(\frac{(1- \lambda_\pm)^{2}}{(1-\lambda_\pm)^{2} + m}\right) \right).
	\end{equation}
	Given the following identities hold:
	\begin{subequations}
	\begin{align}
		\prod_{\pm} (1 - \lambda_\pm) & = -m, \\
		\prod_{\pm} ((1 - \lambda_\pm)^{2} + m) & = m\Delta_{\lambda}^{2},
	\end{align}
	\end{subequations}
	we see that 
	\begin{equation}
	\sum_{\pm} e^{\mp i{\Delta}_{\lambda}t/2} 
		\left(\frac{(1 - \lambda_\pm)^{2}}{(1 - \lambda_\pm)^{2} + m}\right)
	= \cos\left(\frac{t}{2}\Delta_{\lambda}\right) - 
		\frac{(m + \lambda - 1)}{\Delta_{\lambda}} i\sin\left(\frac{t}{2}\Delta_{\lambda}\right).
	\end{equation}
	This proves the claim.
\end{proof}
\end{prop}

\ignore{
The appearance of the exponentials $\exp(-it\lambda/2)$ in \eqref{eq: Lmatrixelement} suggests that 
perfect state transfer in $G$ allows us to construct $G \corona \hseq$ with pretty good state transfer 
by using Kronecker approximation to control all other terms in the sum, reducing \eqref{eq: Lmatrixelement} to a multiple of
$\ebra{u}\exp(-i(t/2)L(G))\eket{v}$. However, applying this heuristic requires some control on the size of the graphs in $\hseq$.
}

The following theorem shows that if $G$ has perfect state transfer, then 
$G \corona \hseq$ has pretty good state transfer for infinite families of graphs $\hseq$.

\begin{thm} \label{thm: pgst-if-pst}
	Let $G$ be a graph on $n$ vertices and $\hseq = (H_1, \dots, H_n)$ be an $n$-tuple of graphs on $m\geq 1$ vertices. 
	Suppose $G$ has perfect state transfer between vertices $u$ and $v$, 
	and let $2^r$ be the greatest power of two dividing each element of the eigenvalue support of $u$. 
	If $2^{r + 1}$ divides $m+1$, then 
	there is pretty good state transfer between vertices $(u,0)$ and $(v,0)$ in $G \corona \hseq$.
	\begin{proof}
		Let $S$ be the eigenvalue support of $u$ in $G$. 
		By Theorem \ref{thm: coutinho-pst-characterization}, we know that the eigenvalues in $S$ are integers.
		Further, if $g$ is the greatest common divisor of all eigenvalues in the support of $u$, 
		then perfect state transfer occurs at times that are odd multiples of $\pi/g$. Since $g$ is an odd multiple of $2^r$,
		this implies that there is perfect state transfer in $G$ at time $\pi/2^r$.
		For any integer $\ell$, we consider the transition element at times
		\begin{equation} \label{eqn: lap-pgst-times}
		t = (4\ell + 2^{1 - r})\pi. 
		\end{equation}
		From Proposition \ref{prop: lap-transition-element}, 
		recall that $\Delta_{\lambda} = \sqrt{(m+\lambda-1)^{2}+4m}$ for each eigenvalue $\lambda$ of $G$.
		We show that for times of the form \eqref{eqn: lap-pgst-times} for specific choices of $\ell$, 
		we have 
		\begin{equation} \label{eq: pgst-if-pst-cosines}
			\cos(\Delta_\lambda t /2) \approx 1
		\end{equation}
		for all eigenvalues $\lambda$ in the support of $u$. 
		Since $2^{r + 1}$ divides $m+1$, we observe that 
		\begin{equation} \label{eqn: zero-is-good}
			\exp\left (-i t (m+1)/2\right) = 1.
		\end{equation}
		If \eqref{eq: pgst-if-pst-cosines} holds, then by Proposition \ref{prop: lap-transition-element}, at time $t$,
		we have
		\begin{equation} \label{eq: pgst-if-pst-matrix-element}
			\ebratuple{u,0} e^{-itL(G \corona \hseq)} \ekettuple{v,0} \approx
			\sum_{\lambda \in \Sp(G)} \exp(-it\lambda/2) \ebra{u} F_\lambda(G) \eket{v} = \ebra{u} e^{-i(t/2)L(G)}\eket{v}.
		\end{equation}
		Since $t/2 \equiv \pi/2^r \pmod{2\pi}$, there is perfect state transfer in $G$ 
		between $u$ and $v$ at time $t/2$, which shows that
		\eqref{eq: pgst-if-pst-cosines} is sufficient for pretty good state transfer in $G \corona \hseq$.

		Note that $\Delta_\lambda$ squares to an integer if $\lambda$ is an integer.
		For each $\lambda$ in the support of $u$, 
		let $c_\lambda$ be the square-free part of $\Delta_\lambda^2$; 
		then $\Delta_\lambda = s_\lambda \sqrt{c_\lambda}$ for some integer $s_\lambda$. 
		From the proof of Theorem \ref{thm: nolaplacianPST} we may see that if $m \ge 1$ 
		and $\lambda$ is a positive integer, then $\Delta_\lambda$ is irrational; this implies that $c_\lambda > 1$	
		if $\lambda > 0$.
		By Lemma \ref{lem: square-free-root-independence}, the disjoint union 
		\begin{equation}
			\{1\} \cup \left\{\sqrt{c_\lambda} : \lambda \in S, \lambda > 0\right\}
		\end{equation}
		is linearly independent over $\mathbb Q$. 
		By Kronecker's Theorem, we may pick integers $\ell, q_\lambda$ such that
		\begin{equation} \label{eq: pgst-if-pst-kronecker}
			\ell\sqrt{c_\lambda} - q_\lambda \approx -\frac{\sqrt{c_\lambda}}{2^{r + 1}}.
		\end{equation}
		If $c_\lambda = c_\mu$ for two distinct eigenvalues $\lambda$ and $\mu$ in the support of $u$, 
		then $q_\lambda = q_\mu$.		
		Multiplying both sides of \eqref{eq: pgst-if-pst-kronecker} by $4s_\lambda$ yields 
		\begin{equation}
			(4\ell + 2^{1 - r}) \Delta_\lambda \approx 4 s_\lambda q_\lambda.
		\end{equation}
		Therefore, at $t = (4\ell + 2^{1 - r})\pi$, we have $\cos(\Delta_\lambda t/2) \approx 1$ for
		$\lambda > 0$.
		To take care of when $\lambda = 0$, note that $\cos(\Delta_0t/2) = \cos((m+1)t/2) = 1$
		from \eqref{eqn: zero-is-good}.
	\end{proof}
\end{thm}

Theorem \ref{thm: pgst-if-pst} provides sufficient conditions for the existence of pretty good state transfer in $G \corona \hseq$ 
for many sequences $\hseq$ of families of graphs. 
For example, consider the $d$-cube $Q_d$. 
The $d$-cube has perfect state transfer at time $\pi/2$ (see Christandl \etal \cite{cdel04}).
Thus, if $m \equiv 3 \pmod{4}$, then $Q_d \corona \hseq$ has pretty good state transfer.
This provides a partial continuous-time analog to the results of Makmal \etal for discrete-time quantum walks on $Q_d \corona H$ \cite{mzmtb14}.

\begin{figure}[H]
	\centering
	\begin{tikzpicture}
		\draw (0:1) -- (90:1) -- (180:1) -- (270:1) -- (0:1);

		\filldraw (90:1) -- ++(-0.5,0.5) circle[radius=0.1];
		\filldraw (90:1) -- ++(0,0.75) circle[radius=0.1];
		\filldraw (90:1) -- ++(0.5,0.5) circle[radius=0.1];
		
		\filldraw (270:1) -- ++(-0.5,-0.5) circle[radius=0.1];
		\filldraw (270:1) -- ++(0,-0.75) circle[radius=0.1];
		\filldraw (270:1) -- ++(0.5,-0.5) circle[radius=0.1];
		\draw (270:1) ++(-0.5,-0.5) -- ++(0.5,-0.25) -- ++(0.5,0.25) -- ++(-1,0);
		
		\filldraw (180:1) -- ++(-0.5,0.5) circle[radius=0.1];
		\filldraw (180:1) -- ++(-0.75,0) circle[radius=0.1];
		\filldraw (180:1) -- ++(-0.5,-0.5) circle[radius=0.1];
		\draw (180:1) ++(-0.5,0.5) -- ++(-0.25,-0.5) -- ++(0.25,-0.5);
		
		\filldraw (0:1) -- ++(0.5,0.5) circle[radius=0.1];
		\filldraw (0:1) -- ++(0.75,0) circle[radius=0.1];
		\filldraw (0:1) -- ++(0.5,-0.5) circle[radius=0.1];
		\draw (0:1) ++(0.5,0.5) -- ++(0,-1);
		
		\filldraw[fill=white] (0:1) circle[radius=0.1] (180:1) circle[radius=0.1];
		\filldraw[fill=gray] (90:1) circle[radius=0.1] (270:1) circle[radius=0.1];
	\end{tikzpicture}
	\caption{The graph $G \corona \protect \hseq$ where $\protect \hseq$ is the sequence of
	non-isomorphic graphs on 3 vertices has pretty good state transfer between the white vertices and
	between the grey vertices.}
\end{figure}
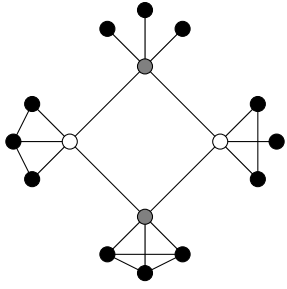

In what follows, we show a different way in which Kronecker's Theorem may be applied in conjunction with Proposition \ref{prop: lap-transition-element}. 

\begin{thm} \label{thm: k2pgst}
	Let $\hseq = (H_1,H_2)$ be a pair of graphs on $m \geq 1$ vertices.
	Then $K_2 \corona \hseq$ has pretty good state transfer between the vertices of $K_2$.
	\begin{proof}
		Let $u$ and $v$ denote the vertices of $K_2$. The Laplacian of $K_2$ has eigenvalues $0$ and $2$, 
		and the corresponding eigenprojectors satisfy
		$\ebra{u} \eigenL{0}{K_2}\eket{v} = 1/2$ and $\ebra{u}\eigenL{2}{K_2}\eket{v} = -1/2$.
		By Proposition \ref{prop: lap-transition-element}, 
		it is sufficient to approximate the following system of equations:
		\begin{align}
			e^{-it(m+3)/2}\cos\left(\frac{t}{2}\sqrt{(m + 1)^2 + 
		            	4m}\right)\approx -1, \\ 
			e^{-it(m+1)/2}\cos\left(\frac{t}{2}(m+1)\right) \approx 1.
		\end{align}
		If $t = 4\pi \ell$ for some integer $\ell$, we note that 
		\begin{equation}
			\exp(-it(m+3)/2) = \exp(-it(m+1)/2) = \cos(t(m+1)/2) = 1.
		\end{equation}
		Thus, it suffices to show that for $t = 4\pi\ell$, we have
		\begin{equation}
			\cos\left(\frac{t}{2}\sqrt{(m+1)^{2} + 4m}\right) \approx -1.
		\end{equation}
		By the proof of Theorem \ref{thm: nolaplacianPST}, we note that $\Delta_2 = \sqrt{(m+1)^2 + 4m}$ 
		is irrational for any positive integer $m$. 
		By Kronecker's Theorem, we may find integers $\ell$ and $s$ such that
		\begin{equation}
			\Delta_2\ell - s \approx 1/2.
		\end{equation}
		This implies that $2\pi \ell\Delta_2  \approx 2\pi s + \pi$, so $t = 4\pi\ell$ gives
		$\cos(t\Delta_2/2) \approx -1$ as desired.
	\end{proof}
\end{thm}

Fan and Godsil \cite{fg13} investigated pretty good state transfer on the double stars $K_2 \corona \emptyg{m}$ 
relative to the adjacency matrix.
They proved that pretty good state transfer relative to the adjacency occured if and only if $1 + 4m$ is not a perfect square.
However, Theorem \ref{thm: k2pgst} shows that pretty good state transfer on double stars 
relative to the Laplacian occurs for all $m$, independent of any number-theoretic conditions on $m$. 
Theorem \ref{thm: k2pgst} also shows that $P_4 = K_2 \corona K_1$ has Laplacian pretty good state transfer,
which was originally observed by Godsil.

\newcommand*{\dsRadius}{1}
\newcommand*{\dsVertices}{6}
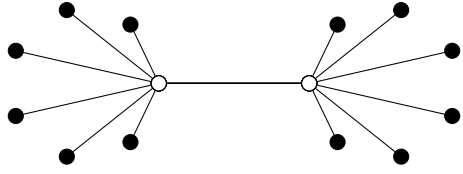
\begin{figure}[h]
	\centering
	\begin{tikzpicture}
		\foreach \s in {0,180} 
		{
			\pgfmathsetmacro{\dv}{\dsVertices + 1}
			\foreach \t in {1,...,\dv}
			{
				\pgfmathsetmacro{\angle}{360*\t/(\dv) + 180}
				\filldraw {(\s:2) + (\angle + \s: \dsRadius)} circle[radius =0.1];
				\draw (\s:2-\dsRadius) -- ($(\s:2) + (\angle + \s: \dsRadius)$);
				\draw (0,0) -- (\s:2-\dsRadius);
				\filldraw[color=black, fill=white] (\s:2-\dsRadius) circle[radius=0.1];
			}
		}
	\end{tikzpicture}
	\caption{$K_2 \corona \emptyg{\dsVertices}$ has pretty good state transfer relative to the Laplacian
			(between the white vertices), 
			but not relative to the adjacency matrix \cite{fg13}.}
\end{figure}

The following example shows that state transfer in $G$ is not necessary for $G \corona \hseq$ to have 
pretty good state transfer.
Coutinho \etal \cite{cggv15} proved that the only distance-regular graph of diameter 2 which admits perfect state transfer
is $\overline{nK_{2}}$ for even $n$.
We show that $\overline{nK}_2 \corona K_1$ has pretty good state transfer for any $n \geq 2$. 
First, we state a useful result about distance-regular graphs.

\begin{lem}[Coutinho \etal \cite{cggv15}, Lemma 4.4] \label{lem: alternating}
Let $G$ be a distance-regular graph with diameter $d$. Suppose $G$ is antipodal with classes of size two and
let $\theta_{0} > \ldots > \theta_{d}$ be the distinct eigenvalues of $A(G)$ with corresponding eigenprojectors
$E_{0},\ldots,E_{d}$. Then, for each $j=0,\ldots,d$, we have
\begin{equation}
A_{d}E_{j} = (-1)^{j}E_{j}.
\end{equation}
Here, $A_{d}$ is the adjacency matrix of a graph obtained from $G$ by connecting vertices $u$ and $v$
if and only if they are at distance $d$.
\end{lem}

\begin{thm} \label{thm: odd-cocktail-party}
	Let $G = \overline{nK_2}$ be the cocktail party graph on $2n$ vertices 
	for some positive integer $n\geq 2$.
	Suppose that $u$ and $v$ are antipodal vertices of $G$.
	Then there is pretty good state transfer in $G \corona K_1$ between $(u,0)$ and $(v,0)$.
	\begin{proof}
		The cocktail party graph is a distance-regular graph with diameter two 
		and is antipodal with classes of size two. 
		Let the eigenprojectors of $A(G)$ be $E_{0}$, $E_{1}$ and $E_{2}$.
		Since $G$ is regular, these are also the eigenprojectors of the Laplacian $L(G)$.
		The corresponding Laplacian eigenvalues are $\lambda_0 = 0$, $\lambda_1 = 2n-2$, and $\lambda_2 = 2n$.

		Let $A_2$ be the adjacency matrix of the graph where $u$ and $v$ are adjacent if and only if they are 
		antipodal in $G$. If $u$ and $v$ are antipodal vertices of $G$, then by Lemma \ref{lem: alternating},
		\begin{equation}
			\eigenA{j} = (-1)^{j}A_2\eigenA{j}
		\end{equation}
		for $j=0,1,2$, which implies that
		\begin{equation}
			\ebra{u}\eigenA{j}\eket{v} = (-1)^j \ebra{u}\eigenA{j}\eket{u}.
		\end{equation}
		Using \eqref{eq: Lmatrixelement} in Proposition \ref{prop: lap-transition-element}, 
		we will pick times $t$ such that $e^{-it\lambda_j/2} = 1$ and
		\begin{equation} \label{eq: cocktail-cosines}
			\cos\left(\Delta_{\lambda_j} t /2\right) \approx (-1)^j,
		\end{equation}	
		where $\Delta_{\lambda_j} = \sqrt{\lambda_{j}^{2}+4}$.
		Note by letting $t = 4\pi \ell$, we have $e^{-it\lambda_j/2} = 1$ for all eigenvalues $\lambda_j$
		and $\cos(\Delta_{\lambda_0}t /2) = \cos(t) = 1$. 
		Now $\Delta_{\lambda_1} = 2\sqrt{1 + (n-1)^2}$ and $\Delta_{\lambda_2} = 2\sqrt{1 + n^2}$. 
		Neither of these are integers for $n \geq 2$.
		Since one of $1 + n^2$ and $1 + (n-1)^2$ is congruent to 2 mod 4 while the other is congruent to 1 mod 4,
		their square-free parts are distinct. Thus, by 
		Lemma \ref{lem: square-free-root-independence}, the set $\{1,\Delta_{\lambda_1}, \Delta_{\lambda_2}\}$ 
		is linearly independent over $\mathbb Q$. By Kronecker's Theorem, there exist integers $\ell, q_1, q_2$ such that
		\begin{subequations}\begin{align}
			\ell \Delta_{\lambda_1} - q_1 & \ \approx \ 1/2, \\
			\ell \Delta_{\lambda_2} - q_2 & \ \approx \ 0.
		\end{align}\end{subequations}
		This implies that, at $t = 4\pi\ell$, equation \eqref{eq: cocktail-cosines} is satisfied.	
	\end{proof}
\end{thm}

\newcommand*{\innerR}{1}
\newcommand*{\outerR}{3/2}
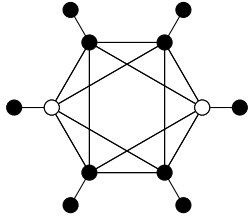
\begin{figure}[h]
	\centering
	\begin{tikzpicture}
		\foreach \s in {0, 60, ..., 359} {
			\filldraw (\s: \innerR) circle[radius = 0.1cm];
			\filldraw (\s: \outerR) circle[radius = 0.1cm];	
			\draw (\s: \innerR) -- (\s:\outerR);
			\foreach \t in {60, 120, 240, 300} {
				\draw (\s: \innerR) -- (\s + \t: \innerR);
			}
		}
		\foreach \s in {0, 180} {
			\filldraw[fill=white] (\s: \innerR) circle[radius = 0.1cm];
		}
	\end{tikzpicture}
	\caption{The graph $\overline{3K}_2 \corona K_1$ has pretty good state transfer
		(between the white vertices), 
		while $\overline{3K}_2$ does not have perfect state transfer.}
\end{figure}

\section{Acknowledgments}

We thank Chris Godsil for valuable discussions about state transfer relative to the Laplacian.
The research of E.A., Z.B., J.M., and C.T. was supported by NSF grant DMS-1262737
and NSA grant H98230-14-1-0141.

\bibliographystyle{plain}

\begin{thebibliography}{10}

\bibitem{normalized14}
{R.} Alvir, {S.} Dever, {B.} Lovitz, {J.} Myer, {C.} Tamon, {Y.} Xu, and {H.}
  Zhan.
\newblock Perfect state transfer in {L}aplacian quantum walk.
\newblock {\em arXiv:1409.5840}, 2014.

\bibitem{bps07}
{S.} Barik, {S.} Pati, and {B.} Sarma.
\newblock The spectrum of the corona of two graphs.
\newblock {\em SIAM Journal on Discrete Mathematics}, 21:47--56, 2007.

\bibitem{b03}
{S.} Bose.
\newblock Quantum communication through an unmodulated spin chain.
\newblock {\em Physical Review Letters}, 91(20):207901, 2003.

\bibitem{bcms09}
{S.} Bose, {A.} Casaccino, {S.} Mancini, and {S.} Severini.
\newblock Communication in {XYZ} all-to-all quantum networks with a missing
  link.
\newblock {\em International Journal of Quantum Information}, 7:713--723, 2009.

\bibitem{cddekl05}
{M.} Christandl, {N.} Datta, {T.} Dorlas, {A.} Ekert, {A.} Kay, and {A.}
  Landahl.
\newblock Perfect transfer of arbitrary states in quantum spin networks.
\newblock {\em Physical Review A}, 71:032312, 2005.

\bibitem{cdel04}
{M.} Christandl, {N.} Datta, {A.} Ekert, and {A.} Landahl.
\newblock Perfect state transfer in quantum spin networks.
\newblock {\em Physical Review Letters}, 92:187902, 2004.

\bibitem{c14}
{G.} Coutinho.
\newblock {\em Quantum State Transfer in Graphs}.
\newblock PhD thesis, University of Waterloo, 2014.

\bibitem{cggv15}
{G.} Coutinho, {C.} Godsil, {K.} Guo, and {F.} Vanhove.
\newblock Perfect state transfer on distance-regular graphs and association
  schemes.
\newblock {\em Linear Algebra and Its Applications}, 478:108--130, 2015.

\bibitem{cl14}
{G.} Coutinho and {H.} Liu.
\newblock No {L}aplacian perfect state transfer in trees.
\newblock {\em arXiv:1408.2935}, 2014.

\bibitem{fg13}
{X.} Fan and {C.} Godsil.
\newblock Pretty good state transfer on double stars.
\newblock {\em Linear Algebra and Its Applications}, 438:2346--2358, 2013.

\bibitem{fg98}
{E.} Farhi and {S.} Gutmann.
\newblock Quantum computation and decision trees.
\newblock {\em Physical Review A}, 58:915--928, 1998.

\bibitem{fh70}
{R.} Frucht and {F.} Harary.
\newblock On the corona of two graphs.
\newblock {\em Aequationes Mathematicae}, 4(3):322--325, 1970.

\bibitem{g11-survey}
{C.} Godsil.
\newblock State transfer on graphs.
\newblock {\em Discrete Mathematics}, 312(1):129--147, 2011.

\bibitem{g12}
{C.} Godsil.
\newblock When can perfect state transfer occur?
\newblock {\em Electronic Journal of Linear Algebra}, 23:877--890, 2012.

\bibitem{gkss12}
{C.} Godsil, {S.} Kirkland, {S.} Severini, and {J.} Smith.
\newblock Number-theoretic nature of communication in quantum spin systems.
\newblock {\em Physical Review Letters}, 109(5):050502, 2012.

\bibitem{gr01}
{C.} Godsil and {G.} Royle.
\newblock {\em Algebraic Graph Theory}.
\newblock Springer, 2001.

\bibitem{hw00}
{G. H.} Hardy and {E. M.} Wright.
\newblock {\em An Introduction to the Theory of Numbers}.
\newblock Oxford University Press, fifth edition, 2000.

\bibitem{mzmtb14}
{A.} Makmal, {M.} Zhu, {D.} Manzano, {M.} Tiersch, and {H. J.} Briegel.
\newblock Quantum walks on embedded hypercubes.
\newblock {\em Physical Review A}, 90:022314, 2014.

\bibitem{r74}
{I.} Richards.
\newblock An application of {G}alois theory to elementary arithmetic.
\newblock {\em Advances in Mathematics}, 13(3):268--273, 1974.

\end{thebibliography}

\end{document}